# Leveraging Cluster Analysis to Understand Educational Game Player Experiences and Support Design

Luke Swanson, Field Day Lab, University of Wisconsin-Madison
David Gagnon, Field Day Lab, University of Wisconsin-Madison
Jennifer Scianna, Field Day Lab, University of Wisconsin-Madison
John McCloskey, Field Day Lab, University of Wisconsin-Madison
Nicholas Spevacek, Field Day Lab, University of Wisconsin-Madison
Stefan Slater, Graduate School of Education, University of Pennsylvania
Erik Harpstead, Human-Computer Interaction Institute, Carnegie Mellon University

**Abstract:** The ability for an educational game designer to understand their audience's play styles and resulting experience is an essential tool for improving their game's design. As a game is subjected to large-scale player testing, the designers require inexpensive, automated methods for categorizing patterns of player-game interactions. In this paper we present a simple, reusable process using best practices for data clustering, feasible for use within a small educational game studio. We utilize the method to analyze a real-time strategy game, processing game telemetry data to determine categories of players based on their in-game actions, the feedback they received, and their progress through the game. An interpretive analysis of these clusters results in actionable insights for the game's designers.

**Introduction**
Playtesting is a well-adopted method for iteratively testing and improving educational games. As a game moves through development phases, members of the target audience are given versions of the game to play, and in exchange generate feedback. This feedback can then be used to validate the design decisions made during the game's development, and to direct the next iterations of work.

Feedback data is commonly captured through qualitative approaches such as direct observation, think-alouds and structured interviews. These approaches generate rich, detailed data that can help game designers understand how their design decisions play out in practice. Design researchers can hear why the player is struggling at a given point, and what approaches they are taking. Further, researchers can potentially observe interactions outside the player's awareness. However, it is difficult to scale these approaches as game development progresses and the testing audience grows. Game data analytics, on the other hand, excel in determining how players interact with the system at scale. Common metrics such as average session duration, number of sessions per user, or max "level" achieved are easily derived. After an up-front cost to implement data logging within the game, analytic approaches scale well to large numbers of players with little additional expense to the design researcher. They are useful for finding technical issues, such as levels in the game that are too challenging, and for assessing engagement of players across sessions. However, this approach's scalability typically comes at the price of a much lower fidelity in understanding player experiences, compared to the qualitative methods described above.

In the past decade, educational researchers have developed tools to synthesize the advantages of qualitative methods with the affordability of quantitative approaches for use with digital learning media. For example, Baker et al. (2020) developed a rigorous observation protocol called BROMP that allows researchers to code participant affect during an observation session, then synchronize observations with log data to train machine learning models for automated detection of these codes in future sessions. Similarly, Shaffer et al. (2016) have developed the method of Epistemic Network Analysis, which can consume large corpuses of human language gathered from instructional systems and use dimensionality reduction and visualization methods to computationally analyze the data. Both of these methods require initial human coding of players' data in order to tell stories about how those players interact with the game. We present a method that similarly marries quantitative analysis of game data with qualitative insight into game designs, but does not require human labeling of raw input data.

## A Method for Clustering Styles of Gameplay

We will now describe a five-stage method of game data analysis that produces a typology of gameplay styles. This method uses existing data mining literature and best practices to enable researchers and designers to see the primary ways their players move through and interact with a game. The first stages may be considered pre-processing steps, in which we select raw data, aggregate the data into well-defined gameplay features, and clean the data to remove invalid or uninformative play sessions. We then apply clustering techniques to identify groups of similar players across three categories of metrics, and finally generate interpretable visualizations to represent the characteristics of each cluster. Our work leverages an existing system for logging, storing and processing educational game telemetry data called Open Game Data (OGD; 2022).

### *Event Selection*

The first step in our method is to identify specific events from the game log data that are most likely to illuminate important aspects of the player experience. Our approach to selecting log data events is informed by previous data mining efforts in educational games (DiCerbo & Kidwai, 2013; Salen & Zimmerman, 2003) and digital assessments (R. S. J. D. Baker & Clarke-Midura, 2013), as well as Owen & Baker's Integrated Design of Event-stream Features for Analysis framework (Owen & Baker, 2020). We choose log data events that fall into three categories:

**Player actions** describe direct player activity in the game space (Salen & Zimmerman, 2003). For example, these events include the player using a tool on an object, placing a resource, pressing a button, or moving a character.
**System feedback** events describe the immediate response by the game system to player actions. This may include audio, visual, or haptic elements, such as points being visually awarded on screen or a sound signaling the successful capture of a resource.
**Progress events** describe the significant movement of the player through the game's designed experience. Examples of progression events include completing a game's level or unlocking a new skill. Different types of progression may take place in parallel; we may refer to each as an axis of progression. For any game, real-time length of play can be used as a progression axis.

Collectively, these events serve to show how the player traversed the game space. After segmenting game events into their respective categories, we choose a small number of events in each category that appear to provide the most fundamental perspectives on the player and game. We base these choices on both the game's design and initial observations of players' experiences.

### *Feature Engineering*

The next step is to aggregate the events selected above into simple numeric values that describe either a single gameplay session or portion of a single session. For example, player actions taken from a level-based puzzle game may lead to features named "pieces moved in level 1", or "total moves in game session". For a game that does not have explicit segmentation, we may arbitrarily divide the session into time-based windows and calculate features such as "moves in 1st 5 minutes"or "total moves in game session". Note that there are many ways to aggregate events. For example,  "level completed" events could be aggregated as "number of levels completed", "percent of levels completed", or "ratio of levels completed to levels started"; all are reasonable options. In general, we prefer simple counts to complicated composite features for this method of analysis.

### *Data Preparation and Cleaning*

Before we move to the data analysis steps, we clean the data through normalization and filtering. The objective is to understand the types of play that occur regularly within a game, so we attempt to include only sessions that represent typical players. We first remove any sessions that did not last long enough to constitute a legitimate play session, as well as excessively long sessions that may result from a player not quitting the game when finished. The thresholds for session length must be determined through analysis of a game's design; a puzzle game may be designed for short 5-10 minute sessions, while designers might intend a story-focused game to be played for 30 minutes at a time. We further omit any outlier sessions, which we define as a session where one or more feature values are at least 3 standard deviations away from the mean.

As a final step before the clustering analysis, we apply transformations to the feature data, to obtain a form that is conducive to clustering. A log transform is applied to any feature that is heavily right tailed (that is, any feature with a few extremely large values). We then center and scale the data. Finally, we apply a dimensionality reduction to the data with the Principal Component Analysis (PCA) algorithm. To pick an appropriate number of output dimensions from PCA, we create a scree-plot (Cattell, 1966) and pick the value at the "knee of the curve."

### Clustering

Standard methods for data clustering find mathematically significant "groups" within the data. Our method uses the k-Means algorithm to cluster the data prepared in the previous step. This algorithm labels each game session as a member of one of $k$ clusters, where $k$ is provided as an input setting for the algorithm. To determine a useful value for $k$, we use the average silhouette score (Rousseeuw, 1987). All else being equal, a higher average silhouette score indicates more distinct clusters. We begin by running the algorithm with $k=2$, then iterate to test higher values of $k$. Typically, the average silhouette score will increase for a few iterations, then begin to decrease. This allows us to quickly find a value of $k$ that maximizes the average score. Once we are satisfied with the choice of $k$, the algorithm generates a dataset where each session has been labeled with a cluster number.

### Visualization and Evaluation

To visualize the results, we use radar charts (Chambers, 1983). The axes of a radar chart are displayed radially. On each axis, a single point is plotted, and these points are connected to define a two-dimensional area. We use the original game data features as axes, rather than the reduced dimensions used for clustering, in order to create charts that are easily interpreted in terms of the game design. There is a question of what value to plot on the chart axes - a cluster is made up of many sessions, so there is no immediately obvious value for each axis. In practice, we found that using a cluster's average value for each feature, represented as a percentage of the overall population average, effectively illustrates the relationship between a cluster and its peers.

Once this process is complete, analysts and designers inspect the radar plots within each category and attempt to describe and name each of the clusters. Good clusters, when charted this way, reveal both categorical differences (seen as different shapes) and differences in magnitude (seen as similar shapes, with differences in their overall area). If the radar plots do not create this effect, then the process should be repeated with new events, new aggregate features, or different parameters for PCA and k-Means.

**Case Study: *Lakeland***

Shown in Figure 1, *Lakeland* (*Lakeland*, 2019) is a systems-based, resource management game that allows students to consider the importance of nutrient cycles in running a self-sustaining, lakeside town. The game was designed for use in secondary science classrooms to address NGSS (CITE) standards related to human impact on the environment, nutrient cycling, and evaluation of complex relationships within ecosystems. From a player perspective, the goal of the game is to "build your city without destroying [your citizens'] lakes". As students engage with the game, they must make decisions that will best assist in expanding their town and building a stable economy, without ruining the environment.

The player generates income for their town by selling produce: corn from crop farms, milk and manure from dairy farms. The money can be used to build more crop and dairy farms, furthering production; to expand population through the addition of homes; or to mitigate algae blooms through the implementation of lake skimming. Players must do this all while balancing the nutrient cycle. In doing so, they learn that manure is a key resource that increases soil fertility to improve crop production, but can also cause algae blooms when it leaches into lakes. Thus, a player's actions and achievements can indicate how well they understand the intricacies of the system (Scianna et al., 2021). There are several avenues to failure in the game: farmers will die if some food is not reserved (or purchased separately) for them to eat, or if they spend too much time in polluted water. Crop farms will fail if the soil nutrition is not supplemented with fertilizer, as each cycle of crop production absorbs nutrients.

**Figure 1**
Screenshot from Lakeland showing an early game state

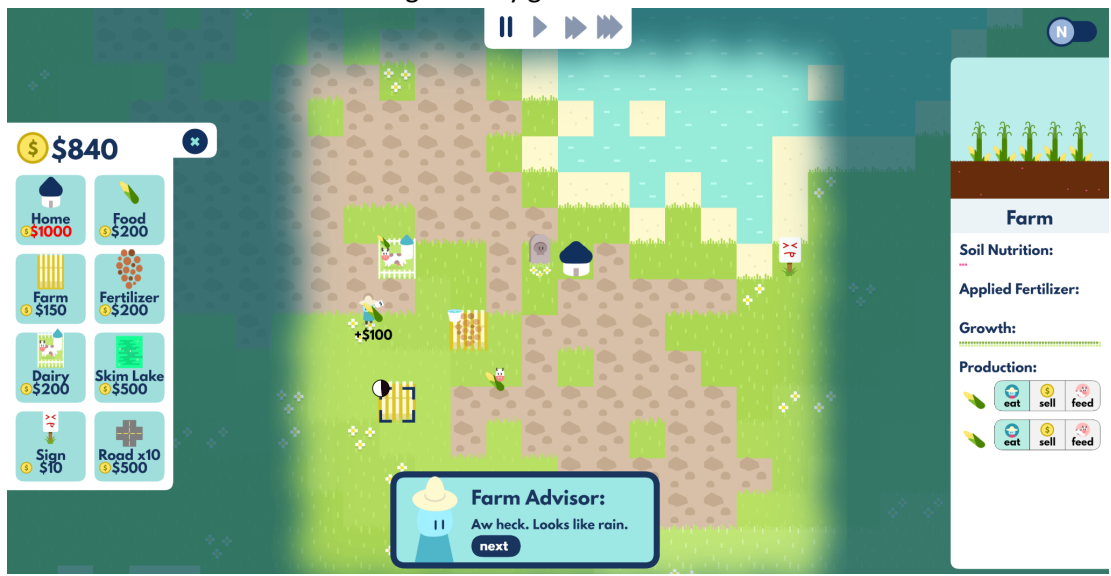

In-game tutorials explicitly scaffold the designed pathway for players in *Lakeland*, nudging players towards self-sustaining strategies. However, the game grants achievements to players in four categories, which may engage different types of players. For example, some players may focus on gaining achievements for a large population, many farms, or money earned. On the other hand, some players may follow a path of destruction, progressing primarily along the algae bloom track. This tension between the designed (and instructed) pathway and the varied achievements leads to opportunities for divergent play. In this case study, the objective is to use our method to evaluate the game's existing design, by examining the players' decisions and progress, and how they interact with game feedback.

**Results**
The data set for this study contained 32,227 anonymous sessions from December 2019 and January 2020. Because *Lakeland* does not have a concept of levels, we divided each gameplay session into five-minute windows, with 30 seconds of overlap between windows. Features using the windowed data were only calculated for the first two windows of each session, in order to standardize the amount of per-player gameplay analyzed.

In the event selection and feature engineering phases, we created five features for each category: For player actions, we count the number of homes, crop farms, and dairy farms purchased, and total purchases overall (which includes buying manure and food directly). The fifth feature counts the number of tiles on which the player "hovered" before placing crop farms - a "hover" shows the player the tile's initial soil nutrition. Our feedback features include the numbers of deaths, crop failures, food (crops) produced, milk produced, and algae blooms. For progression, we have the number of player achievements in the following areas: population size, total money saved, farms built, and bloom count. Finally, we use total real-world play time as a progression feature. We should note also that our progression features used players' full session data, rather than windowed data.

In the data cleaning step, we applied our standard set of filtering and transform rules. We chose 5-45 minutes as our range of "valid" session durations, and additionally filtered out any sessions with fewer than 10 player action events. Outlier removal was performed separately on the *actions* and *feedback* event categories, and not at all for the *progression* category (all progression features had small maximum values, negating the effect of outliers). After filtering, we had 5486, 6448 and 10164 sessions for the *actions*, *feedback* and *progression* categories, respectively. We used PCA to reduce data to 2 dimensions for all categories, with the number of dimensions in each case determined by reading the corresponding scree plots. In the clustering step of the analysis, silhouette scores led us to select 6 k-Means clusters for the *actions* category and 7 k-Means clusters for *feedback* and *progression*. Finally, we generated one set of radar charts for each feature category. Our analysis and interpretation of the charts' "stories" are described next.

**Figure 2**
Player actions clusters. Group names (L-R): Planners, Livestockers, Balanced, Inactives, Vegans, Capitalists

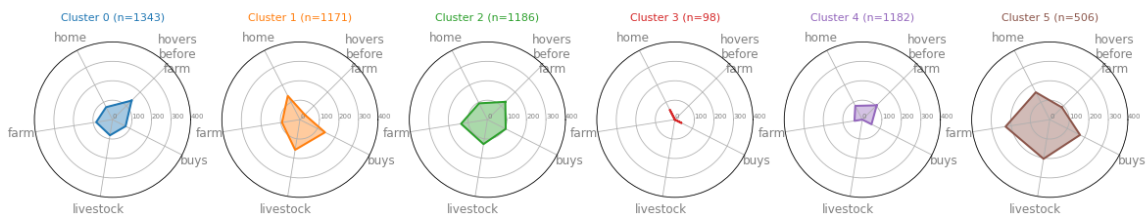

### *Interpretation of Results*

We found 6 player action clusters (see Figure 2) to describe player behavior. Cluster 0 (Planners) and cluster 2 (Balanced) are dilations of one another, where players in both clusters regularly use the hover tool, but *Planners* generally make fewer purchases overall. Cluster 1 (Livestockers) focus on homes and livestock to create a dairy-driven economy and perform less inspection of land conditions before buying. Cluster 3 (Inactives) appear to be the few players (< 2% of sessions) who carried out enough actions to pass the data filtering step, but did little else in the first two windows of play time. Cluster 4 (Vegans) never establish a livestock farm. Cluster 5 (Capitalists) players make the most purchases of any group by far, but are the least common group among "active" players.

**Figure 3**

Feedback clusters. Group names (L-R): Low Feedback, Good Dairy Poor Lakes, Cemetery Town, No Feedback, Prolific Growth, Balanced, Corn-centric

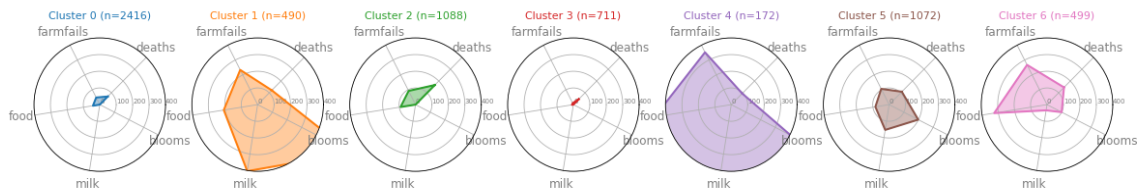

In the Feedback category (see Figure 3), Clusters 0 (Low Feedback) and 3 (No Feedback) have very few feedback events. Together, they represent nearly half of all players. Cluster 1 (Good Dairy Poor Lakes) players saw high yields from dairy farms, but experienced many algae blooms. Players in cluster 4 (Prolific Growth) faced significant farm and ecological failures, but received over 4 times as much food and milk as the average player. In Cluster 2 (Cemetery Town), the players' towns had the greatest number of deaths, and almost no milk production. Cluster 6 (Corn-centric) shows relatively functional communities with more food, crop failures and deaths than most groups, again with almost no milk production.

**Figure 4**
Progression clusters. Group names (L-R): Lackluster, Money-centric, Achieve and Destroy, Food-centric, Caution ahead, Lost in Town, Happily in Harmony

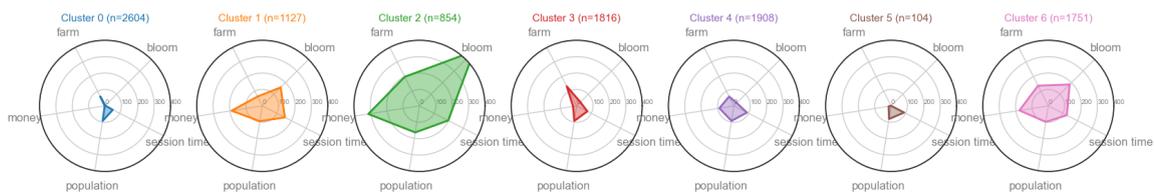

Progression clusters (see Figure 4) vary dramatically across average session time. Cluster 0 (Lackluster), which was the most common group, contained players with the least time played and fewest achievements. In contrast, cluster 1 (Money-centric) played longer than average, and had a high ratio of money achievements to farm achievements. Cluster 2 (Achieve and Destroy) members played longer than any other group, making significant progress on all fronts, but their greatest leap in achievement comes on the bloom axis. Players in clusters 3-5 had average session lengths, with varying degrees of achievement. Cluster 6 (Happily in Harmony) players managed to go above the average on each axis, but did not move as far along the algae bloom axis as their *Achieve and Destroy* peers.

**Discussion**
The player-categorizing clusters help to affirm existing design choices. For example, the analysis of player actions reveals the hovering feature is relatively well used by players, with only the *Livestockers* underutilizing the feature. However, *Livestockers* are also alone in their focus on the building of dairy farms, which are not dependent on soil nutrients and proximity to water like crop farms. Thus, they have less use for the hover tool's features, indicating consistency between the tool design and its use by players. Another group, *Vegans*, avoided livestock altogether, but other clusters typically had average usage of livestock and corn farms, indicates the cost vs. reward structure for the primary game resources is well-balanced. The *Good Dairy Poor Lakes* feedback cluster indicates the game's ecosystem simulation works as intended, as the players' high level of dairy production generates manure that contributes to algae blooms. Further, we can identify a particular group of struggling players in the *Cemetery Town* cluster, where the average food production numbers provide evidence these players experienced growing towns before their farms failed and farmers began to die, affirming some players experience the game's avenues to failure. On the other hand, there are some apparent design weaknesses. Nearly half of the players fell into groups that received below-average amounts of feedback, indicating the game experience may not have been very interactive for these players. The results for the progression category show that only players in the *Money-centric, Achieve and Destroyers,* and *Happily in Harmony* clusters ever had a bloom achievement on average, which is a central component of the game's stated learning goals. Bloom achievements primarily occurred in clusters that had session times above the overall average, suggesting the game should speed the algae bloom process so more players reach this point.

## Conclusion

In this paper, we have described a general, scalable method for identifying patterns of gameplay in a way that serves game designers' need for qualitative insights into player behavior.  We applied this method to an open-ended, achievement-driven game. The framework for dividing events into three categories let us bypass some of the typical guesswork associated with feature engineering. The use of radar plots to visualize clusters was conducive to generating qualitative insights from a quantitative analysis, and understanding the paths players take in exploring the game space. This allowed us to validate some aspects of the game's design, while identifying weaknesses in others. Our method adds a valuable tool to the box of playtesting techniques.

## Acknowledgments

This work was funded by NSF (DRL# 1907384/1907437), the National Institute of Food and Agriculture (# 2017-67003-26055), and the Wisconsin Department of Public Instruction.